# Magnetic Nanoparticle Chains in Gelatin Ferrogels: Bioinspiration from Magnetotactic Bacteria

*Sebastian Sturm, Maria Siglreitmeier, Daniel Wolf, Karin Vogel, Micha Gratz, Damien Faivre, Axel Lubk, Bernd Büchner, Elena V. Sturm (née Rosseeva),\* and Helmut Cölfen\**

Inspired by chains of ferrimagnetic nanocrystals (NCs) in magnetotactic bacteria (MTB), the synthesis and detailed characterization of ferrimagnetic magnetite NC chain-like assemblies is reported. An easy green synthesis route in a thermoreversible gelatin hydrogel matrix is used. The structure of these magnetite chains prepared with and without gelatin is characterized by means of transmission electron microscopy, including electron tomography (ET). These structures indeed bear resemblance to the magnetite assemblies found in MTB, known for their mechanical flexibility and outstanding magnetic properties and known to crystallographically align their magnetite NCs along the strongest <111> magnetization easy axis. Using electron holography (EH) and angular dependent magnetic measurements, the magnetic interaction between the NCs and the generation of a magnetically anisotropic material can be shown. The electro- and magnetostatic modeling demonstrates that in order to precisely determine the magnetization (by means of EH) inside chain-like NCs assemblies, their exact shape, arrangement and stray-fields have to be considered (ideally obtained using ET).

## 1. Introduction

Magnetite ($Fe_3O_4$) is an interesting and widely studied mineral, which is not only found as a geological deposit, but also in living organisms. In magnetotactic bacteria (MTB) for example, ferrimagnetic magnetite nanocrystals (NCs) of high saturation magnetization and with high coercive fields[1] are arranged in intracellular chains. These chains act as sensitive magnetic antenna that enable navigation in the earth's magnetic field.[2] These natural specimens and their synthetic analogues represent an intriguing material for a variety of potential applications, e.g., magnetotransport, magnetic separation, micromechanical sensors, soft actuators as magnetic robotics.[3] However, for any biomedical application, e.g., immune[4] and receptor binding,[5] magnetic cell separation,[6] DNA extraction,[7] magnetic resonance imaging,[8] magnetic hyperthermia therapy[9] etc.,[10] NCs would need additional surface functionalization[2a] and their size must not exceed the superparamagnetic limit to avoid uncontrolled aggregation. Different routes to biomimetic and bioinspired syntheses are known, which offer the opportunity to control particle size, shape, orientation and organization of nanosized material.[11] The formation of ferrimagnetic magnetite (>30 nm) with controllable size and shape generally requires high-temperature synthesis methods, and only a few approaches exist, where control over particle size was shown at ambient conditions using additives.[11a,12]

Instead of imitating the complicated natural growth mechanism in MTB biomimetically, involving cell constituents, we have developed a simpler bioinspired synthesis in a gelatin hydrogel matrix at ambient conditions based on an already successfully established protocol for super-paramagnetic hydrogels,[13] more practical for application. Gelatin is

S. Sturm, Dr. D. Wolf, Dr. A. Lubk, Prof. B. Büchner
Institute for Solid State Research
Leibniz Institute for Solid State and Materials Research Dresden
Helmholtzstraße 20, 01069 Dresden, Germany
Dr. M. Siglreitmeier, Dr. E. V. Sturm (née Rosseeva), Prof. H. Cölfen
Physical Chemistry
Department of Chemistry
University of Konstanz
Universitätsstraße 10, 78457 Konstanz, Germany
E-mail: elena.sturm@uni-konstanz.de; helmut.coelfen@uni-konstanz.de

Dr. K. Vogel
Institute of Resource Ecology
Helmholtz-Zentrum Dresden-Rossendorf
Bautzner Landstraße 400, 01328 Dresden, Germany
M. Gratz
Experimental Physics
Saarland University
Campus D2 2, 66123 Saarbrücken, Germany
Dr. D. Faivre
Department of Biomaterials
Max Planck Institute of Colloids & Interfaces
Science Park Golm, 14424 Potsdam, Germany
Dr. D. Faivre
CEA/CNRS/ Aix-Marseille Université
UMR7265 Institut de biosciences et biotechnologies
Laboratoire de Bioénergétique Cellulaire
13108 Saint Paul lez Durance, France

denatured collagen, which has preserved the capability to form triple helices depending on the molar mass of the gelatin molecules and is forming a thermoreversible gel. This gel contains a mesh, in which nanoparticles like magnetite can be located and can therefore serve as a template structure for the synthesis of magnetite nanoparticles.[13a,14] This is similar and inspired by the magnetosome vesicles in MTB.

## 2. Results and Discussion

Gelatin hydrogel samples in different concentrations (6–18 wt%) are loaded with Fe(II)-ions, where magnetite NC formation is initiated by placing the Fe(II) loaded gels into a solution of $KNO_3$ (0.5 M) and KOH (0.1 M). The gel color change can be followed visually transitioning from a dark green to deep black upon placing the iron-loaded gel into the $NO_3^-$ and $OH^-$ solution. A dark green color possibly indicates an intermediate product of green rust, like in the transformation of various oxyhydroxides and iron oxides to magnetite.[15] After the first reaction cycle (RC), the mineral loaded gels can be reintroduced into the iron(II) solution with following repetition of the magnetite precipitation. The as described RC can be repeated for several times and results in an increasing mineral content of the ferrogels varying from 30 to 50 wt% in the dry material.

**Figure 1**a–c displays transmission electron microscopy (TEM) images of magnetite NCs prepared in a gelatin hydrogel matrix. The bright-field (BF)-TEM images clearly show spherical particles with an average size of 50 nm arranged in chains due to dipolar interactions between the ferrimagnetic particles.[12d,16] The selected area electron diffraction (SAED) pattern (Figure 1d) shows only Bragg reflections corresponding to the magnetite crystal structure, in good agreement with the powder X-ray diffraction (XRD) patterns (Figure S1, Supporting Information).

For the investigated gelatin concentrations (6% to 18%) no changes in size and morphology of the obtained NCs were found. On the first glance, this observation is surprising, since the synthesized particle size is at least 30 nm, which is bigger than the average gelatin hydrogel mesh diameter (22 nm) investigated by small angle neutron scattering (SANS),[13a] meaning that the particles are expanding the gelatin hydrogel structure and/or partially incorporating the gelatin molecules inside the NCs. Thus, the gelatin gel does not only serve as an inert matrix limiting the diffusion process of the reacting ions and slowing down the growth process of magnetite NCs, but also provides active nucleation sites promoting the formation of magnetite NCs.[14] Recently performed molecular dynamic simulations show that a triple helical (Gly-Hyp-Pro)$_n$ peptide favors the binding of the Fe(OH)$_x$ motifs (forming in the early stage of magnetite crystallization) and thus induces the intergrowth of macromolecules and magnetite NCs already at the precursor stage, which is also recently proved by SANS.[13] Furthermore, the high-resolution (HR)-TEM images of isolated magnetite NCs clearly reveal the highly pronounced mosaic-like defected structure, although they demonstrate single-crystal like diffraction properties (Figure S2, Supporting Information). Similar observations were also found for the case of magnetite particles synthesized in presence of other organic macromolecules[12d,17] as well as fluorapatite-gelatin composites crystallized in gelatin gel.[16c,18]

The observed chain like structures bear a close resemblance to those found in MTB. The particles are spatially separated within the chain, retaining the flexibility of the assembly. This property has recently been recognized as essential for the MTB's propulsion mechanism and is therefore of high interest for nanorobotics.[3b] In order to reveal the preferred crystallographic orientation of the NCs within the chains, a systematic analysis of diffractograms retrieved from the HR-TEM images was performed. Figure 1e and Figure S3 (Supporting Information) demonstrate that although the NCs have different projected crystallographic orientation their <111> axis, the easy direction of magnetization in magnetite, is tending to follow the elongation of the chain. Similar behavior has been reported by Simpson et al. analyzing the crystallographic alignment of magnetite NCs within chain-like structures extracted from MTB.[19] Thus, these assemblies show a clear 1D mesocrystalline ordering of NCs.[16c,20]

In order to investigate the magnetic properties of the chains of magnetite NCs, off-axis electron holography (EH) was applied. Using a so-called Lorentz lens allows investigating magnetic samples under virtually magnetic field-free conditions, thus not disturbing the configuration of the particles during measurement. EH permits reconstructing the phase of the electron wave that has passed through the sample. Sensitive not only to electrostatic potentials, i.e., the mean inner potential (MIP) defined as the average Coulomb potential of the crystal, but also to magnetic fields within and around the sample.[21]

Our EH observations on the magnetite chains confirm that the magnetization of the NCs is indeed following the axis of the chain, maximizing the overall magnetic moment. **Figure 2** shows the EH analysis of one exemplary chain composed of 65 nm sized particles. The projected electrostatic potential data (Figure 2a) reveals once again the highly mosaic-like defected structure of the NCs, whereas magnetically they seem to exhibit a single magnetic domain-like behavior. In order to evaluate further the properties of the NCs in terms of composition and internal magnetization $M$, the experimentally obtained data is compared with an electrostatic model (Figure 2b) and its magnetostatic simulation (Figure 2d). To this end, the shape of the chain has been approximated by a model of five touching spheres of 65 nm diameter (Figure 2b); with the MIP $V_{MIP}$ (reflecting the composition) and the magnetization $M$ as free fit parameters. This approach automatically accounts for the demagnetizing fields that reduce the magnitude of the magnetic flux measured in 2D projection by EH. The comparison of measured and simulated profiles (Figure 2e,f) taken normal to the chain's long axes reveals a good agreement between both values of $V_{MIP} = 12.0$ V and $M = 0.27$ T.

In order to investigate the effect of the gelatin on the structural and magnetic properties of magnetite chains, we also examined a nongelatin sample (prepared by the oxidation method in aqueous-solution at low temperature) for comparison. In contrast to our gelatin-bearing specimens the magnetite particles prepared without gelatin are clearly facetted. To reveal the crystallographic shape of the particles, we performed bright-field-ET. As one can see in **Figure 3** and Movie S1 (Supporting Information), the habit of NCs is quite diverse and is composed of {100}, {111}, and {110} faces of different ratio. Furthermore, the particles within the chain are in direct contact making the chain stiff in comparison to the gelatin bearing specimen. Notwithstanding, the particles <111> easy axis still follows the chain elongation (Figure 3b). The apparent slight deviation at the end of the chain is likely a projection artifact, since the end of the chain is pointing slightly upward from the supporting grid, as can be seen in the Movie S1 (Supporting Information).

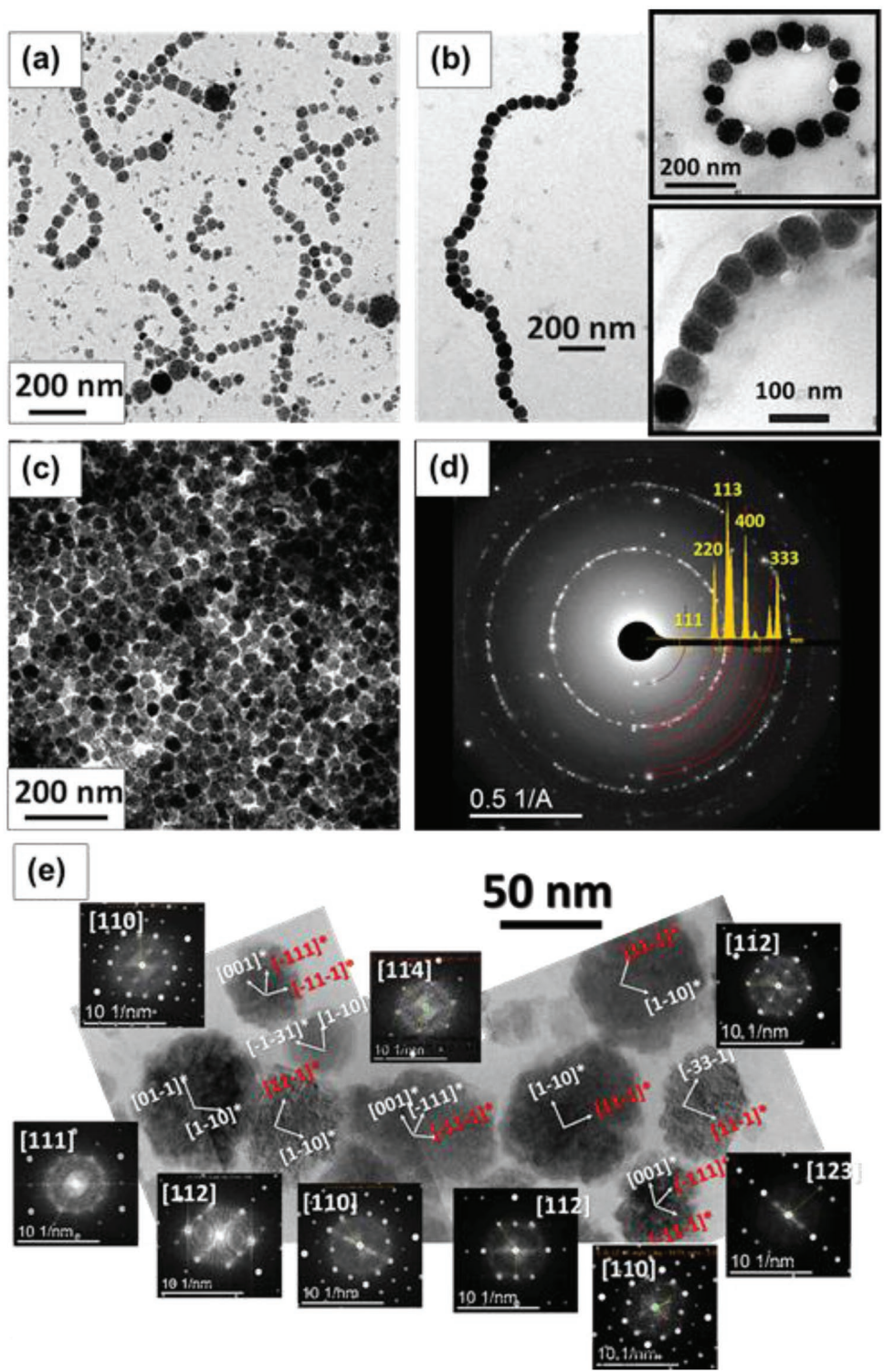

**Figure 1.** Structural and morphological features of magnetite NCs and their chain-like self-assembly in ferrogels prepared with 12 wt% gelatin after 4 RC. a,b) BF-TEM images of magnetite NCs and chains isolated from the ferrogel c) BF-TEM image of a thin section (prepared by means of Ultramicrotomy) of dry ferrogel. d) SAED pattern collected from magnetite particles shown in image a) overlaid by simulated electron diffraction patterns of magnetite (powder, ring pattern). e) HR-TEM image of a magnetite chain, showing arrangement and crystallographic orientation of NCs within the assembly. Diffractograms obtained from each NC using a small ROI mask and overlaid by a simulated electron diffraction pattern of magnetite in the fitted zone axis. The orientation of <111> (corresponding to the strongest magnetization easy axis of magnetite) is indicated in red.

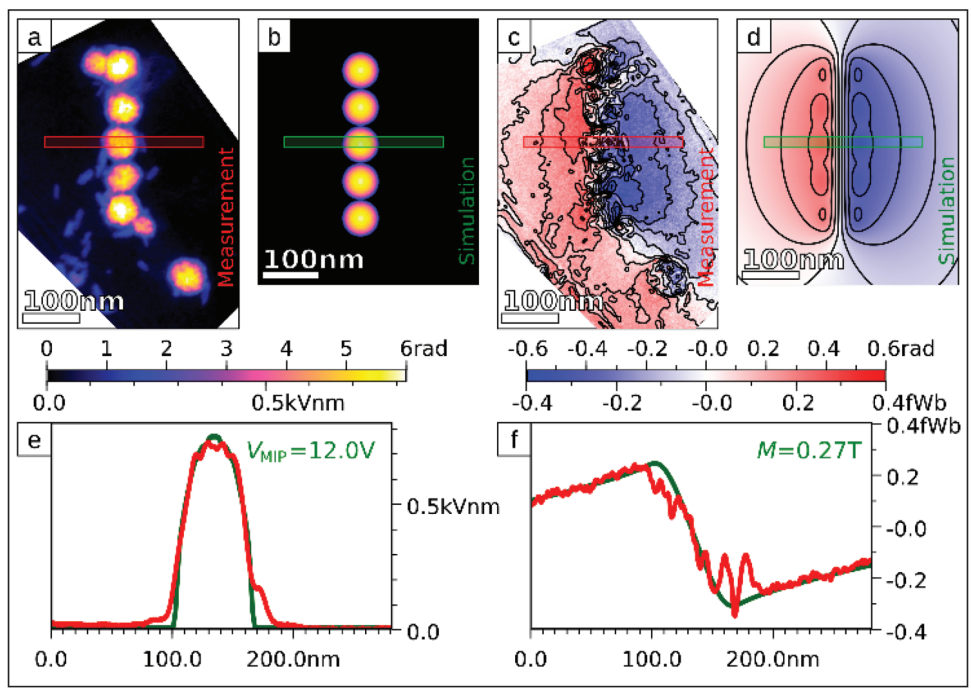

**Figure 2.** EH of a magnetite chain obtained from ferrogels (prepared with 12 wt% gelatin after 4 RCs). a,b) Projected electrostatic potential and c,d) magnetic flux, as measured or obtained by simulation, respectively. Comparison of electrostatic e) and magnetic f) profiles at the position indicated by the red and green bars in (a–d) illustrate the good agreement between experiment and simulation at values of $V_{MIP} = 12$ V and $M = 0.27$ T.

The EH investigation (**Figure 4**) of the same chain yields a MIP of $V_{MIP} = 13.5$ V and a magnetization inside the particles of about $M = 0.28$ T. In order to precisely determine the

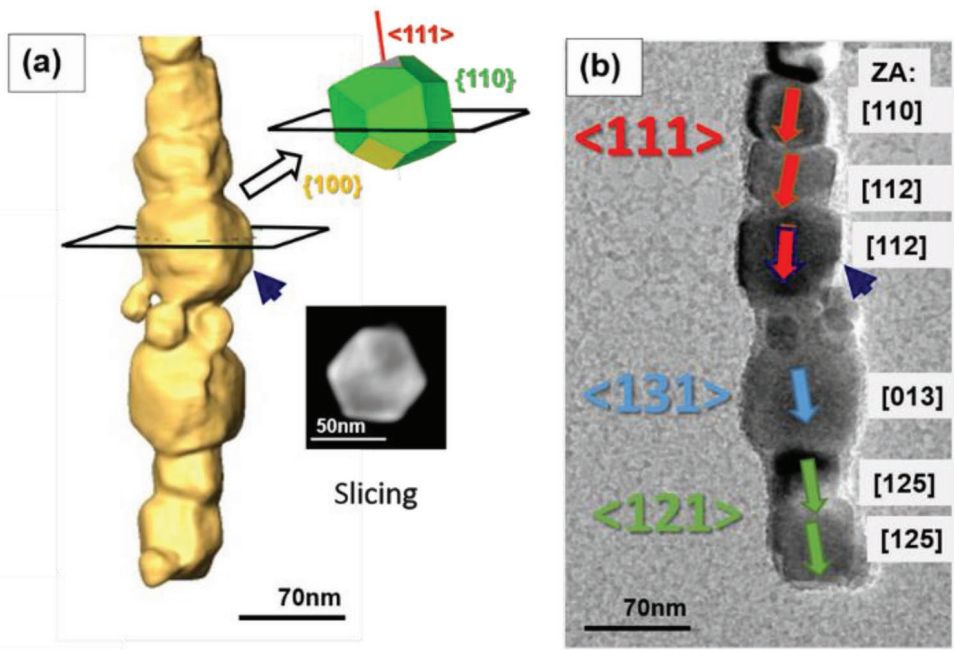

**Figure 3.** a) ET reconstruction of the magnetite chain (synthesized without gelatin) and crystallographic shape (slightly truncated rhombidodecahedra) for one selected NC and b) respective HRTEM image illustrating the crystallographic orientation of magnetite NCs.

1905996 (4 of 8)

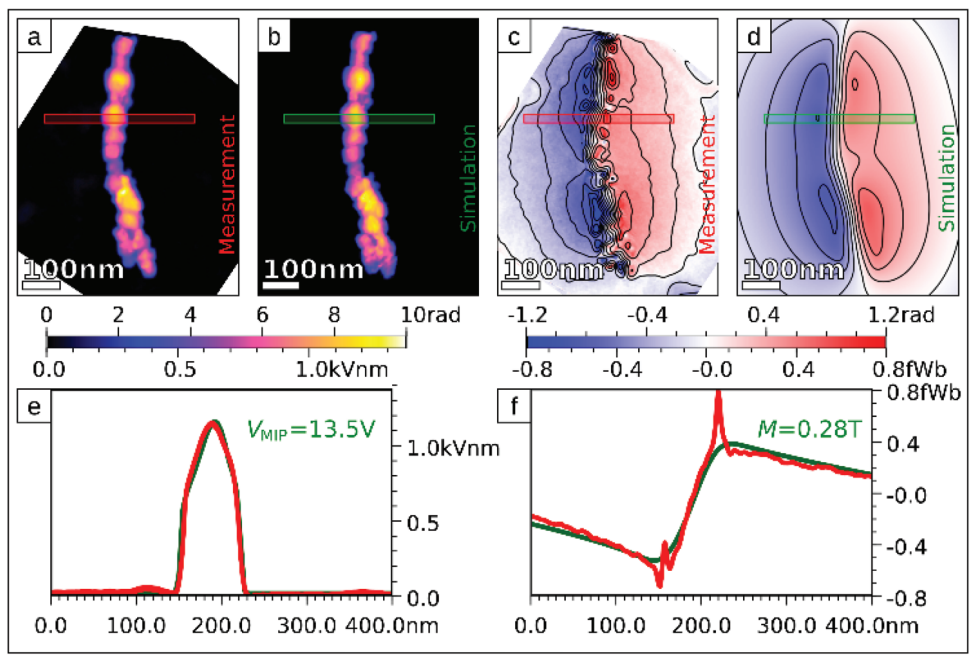

**Figure 4.** EH of a magnetite chain prepared without gelatin. a,b) Projected electrostatic potential and c,d) magnetic flux, as measured or obtained by simulation using the shape obtained from ET (Figure 3), respectively. Comparison of electrostatic e) and magnetic f) profiles at the position indicated by the red and green bars in (a–d) illustrate the good agreement between experiment and simulation at values of $V_{MIP}$ = 13.5 V and $M$ = 0.28 T.

magnetization inside the particles, we used the shape of the chain as obtained by ET (Figure 3) in our simulation. Moreover, we assumed a homogeneous magnetization in one direction throughout the whole chain. The lower MIP of the gelatin-bearing chain compared to the pure magnetite sample strongly suggests the incorporation of gelatin. Considering further that the theoretical MIP (using the ionic potential parameterization of ref. [22]) of bulk magnetite is around 15.7 V, a certain amount of disorder and surface oxidation may be present in both cases. In spite of this change in composition, the magnetization of both chains remains basically the same, with the size of the individual NCs in a similar range. The magnetization of our synthetic magnetite chains prepared with or without gelatin gel is only around 50% of the bulk value for magnetite,[23] most likely because the synthetic crystals are smaller, not perfect single crystals (due to gelatin, defects, etc.), and may contain oxidized surface layers of maghemite with lower magnetization as suggested by the reduced MIP values. In particular the presence of antiphase boundaries has been recently identified as a major source for reduced magnetizations in small magnetite NCs.[24] Indications of reduced magnetization have also been reported recently for a similar system of synthetic magnetite particles (≈40 nm) formed in the presence of polyarginine and also arranged in chain-like structures.[12d]

This shown tendency of the magnetic NCs to align their respective field might be utilized in materials, which can be used for storing magnetic information. In order to examine if magnetic information can be written into the synthesized ferrogel, particle arrangement upon drying under an external magnetic field was investigated. The scanning electron microscopy (SEM) image (**Figure 5**a) reveals an anisotropic inhomogeneous material displaying particles aligned parallel to the applied field.

To get information about the magnetic properties of the material, angular dependent magnetization measurements of hysteresis loops were performed. Plotting the angular dependence of the remanent magnetization normalized by the measured value of the saturation magnetization (MR/MS) against sample orientation (Figure 5b) reveals that the ferrogel magnetization is clearly angle dependent with a maximum value of MR/MS in between 15° and 30° and a minimum in between 90° and 105°. Thus, the orientation of the NCs inside the ferrogel was successful, but not as pronounced as found in MTB, which show ten times higher orientation comparing normalized angular dependent measurements.[25] One possible explanation for this observation could be that the crystallites in this synthesis procedure are no perfect defect-free single crystals and show sphere-like morphology and no shape-anisotropy. The latter would increase the magnetic anisotropy of each single particle as it is observed for the MTB. Additionally, the dense particle packing and therefore agglomeration can cause dipolar interactions between the particles, which can influence the magnetic properties of the material and thus decrease magnetic anisotropy. Apart from that it has to be pointed out that the ferrogel samples with a low magnetite content of 30 wt% show a self-healing ability at elevated temperatures of 30 °C (see Figure S5 in the Supporting Information) which additionally adds interesting properties to the material for potential applications.

## 3. Conclusion

In summary, we have presented easily accessible synthesis methods for ferrimagnetic magnetite nanochains and ferrogels composed of gelatin hydrogel and magnetite NCs.

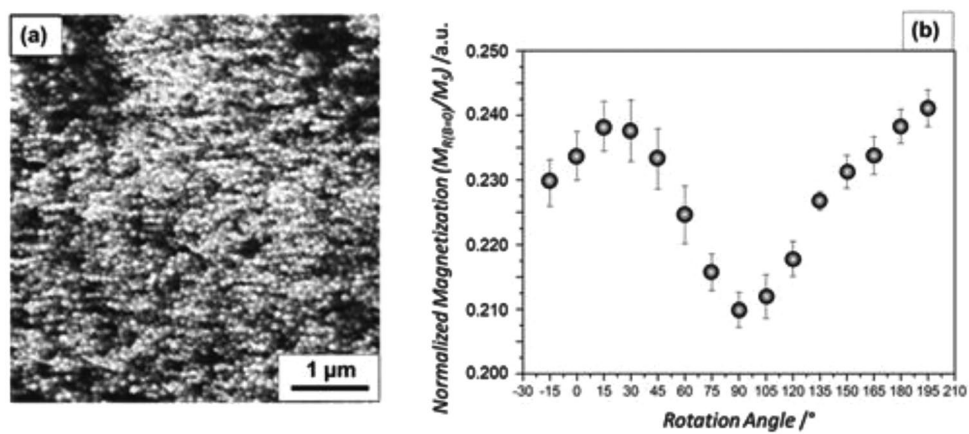

**Figure 5.** a) SEM image of ferrogel surface dried under an external magnetic field of 1 T. b) Angular variations for the hysteresis parameter of the oriented ferrogel plotting the remanence ratio against angular orientation of the sample.

TEM analysis (inc. electron tomography and holography) of the chain like structures of ferrimagnetic NCs prepared with gelatin reveals great resemblance with magnetite from MTBs. Especially the flexibility of these chains compared to their nongelatin counterparts could be a promising playground, e.g., for propulsion mechanisms in nanorobotics. Furthermore, it was shown that the magnetic dipole of these chains inside the ferrogel can be aligned under an external magnetic field creating a magnetically anisotropic material. Last but not least, structural and magnetic properties of magnetite chains synthesised with and without gelatin were compared. The performed electrostatic and magnetostatic modelling not only verified the experimental observations but also provided great insight into fundamental understanding of structural and magnetic properties of magnetite chain-like structures, prepared under these two synthesis conditions. It is also demonstrated that in order to precisely determine via electron holography the magnetization inside particles forming the chain-like assembly, their exact shape and arrangement has to be considered (ideally obtained using a tomographic reconstruction).

## 4. Experimental Section

*Synthesis of Magnetite Nanoparticles and Chains—Materials*: The following commercially available chemicals were purchased and applied in the synthesis without further purification: $FeSO_4·7H_2O$ (Sigma-Aldrich), KOH (Sigma-Aldrich), $KNO_3$ (Sigma-Aldrich), Gelatin Type B (≈225 Bloom, Sigma-Aldrich), 4-chloro-m-cresol (Fluka), Methanol (VWR). For the preparation of the reactant solutions double-distilled and deionized (Milli-Q) water was used. All solutions were degassed with nitrogen before usage.

*Synthesis of Magnetite Nanoparticles and Chains—Gel Synthesis*: The gelatin hydrogels were prepared as already described in ref. [13]. Here briefly, different amounts of gelatin were prepared in disposable base molds in concentrations ranging from 6 to 18 wt%.

*Synthesis of Magnetite Nanoparticles and Chains—Synthesis of the Ferrogel*: In situ mineralization of magnetite nanoparticles in gelatin hydrogel was carried out similar to an already described synthesis procedure. The synthesis is based on the so-called partial oxidation of $FeSO_4$ by $KNO_3$. Each gelatin hydrogel sample was introduced into a solution, containing $FeSO_4$ (0.2 M) where it was left for 96 h at 6 °C. The iron(II)-loaded gels were washed with water and placed in a solution of 0.1 M KOH and 0.5 M KNO3 for at least 1200 min at 6 °C.

*Synthesis of Magnetite Nanoparticles and Chains—Synthesis of Magnetite Nanoparticles without Gelatin*: Magnetite nanoparticles were synthesized following the modified approach developed by Sugimoto and Matijević.[15] First a solution of 20 mL containing $FeSO_4·7H_2O$ (0.3 mol $L^{-1}$) was prepared (the pH was adjusted to 1 by addition of $H_2SO_4$, 0.01 mol $L^{-1}$). 80 or 180 mL of the base KOH (0.1 mol $L^{-1}$) and the oxidation medium $KNO_3$ (0.5 mol $L^{-1}$) were placed in a three-necked-flask. In case of precipitation in the presence of gelatin, the gelatin stock solution was then supplied in appropriate amounts of either 1 or 0.1 mg $mL^{-1}$ and the pH was reset. All stock solutions were degassed with nitrogen prior usage. Magnetite formation was initiated by adding the iron sulfate solution at a dropping rate of 1.25 mL $min^{-1}$ under vigorous mechanical stirring. The whole synthesis procedure was carried out under nitrogen atmosphere. After addition of iron solution, the resulting reactant solution was kept at 6 °C for 24 h. The last step showed the formation of a black precipitate which can be attracted by an external magnetic field. The resulting product was washed magnetically with degassed Milli-Q water until the product suspension showed pH neutrality. The precipitate was separated by centrifugation (7500 rpm, 40 min) washed with ethanol and dried under a weak nitrogen stream for further investigations.

*Analytical Methods—Thermogravimetric Analysis (TGA)*: The mineral content of the hydrogels was determined by means of TGA (Netzsch, Selb, Germany). Measurements were carried out at a heating rate of 5 K $min^{-1}$ under a constant oxygen flow. Samples were scanned from 293 to 1273 K.

*Analytical Methods—X-Ray Diffraction Analysis*: The phase of the iron oxide nanoparticles was characterized by XRD at the µ-Spot synchrotron beamline (BESSY II, Helmholtz Zentrum Berlin) with a 100 µm beam at 15 keV. The measurements were performed by Marc Widdrat in the group of Damien Favre, MPI of colloids and interfaces in Golm.

*Analytical Methods—Magnetic Characterization*: The magnetic properties of the iron(II)-loaded gel were investigated using a vibrating sample magnetometer (VSM 7300, Lakeshore). The sample film was fixed in plane to the end of the sample rod by a capton tape. The magnetic moment as measurement quantity of the sample was measured at room temperature of 21 °C in field direction as function of the applied homogeneous magnetic field H for a variety of different rotational angles. The rotation angle l° between the magnetic field and the induced anisotropy axis was modified in steps of 15° in the range of 0°–180° ± 15° (where H is parallel to the alignment field for values close to 0° and 180° or perpendicular for values around 90°). From the measured hysteresis loops, the angular dependent remanent magnetizations MR normalized by the measured value of the saturation magnetization MS were calculated.

*Analytical Methods—Scanning Electron Microscopy*: For SEM analysis a Zeiss Neon 40 EsB operating in high vacuum was used. An InLens and SE detector was used for signal collection and an acceleration voltage of 2 kV was chosen for recording the images. The specimens were coated with a thin layer of gold in order to avoid charging effects.

*Analytical Methods—Transmission Electron Microscopy*: Medium resolution TEM and HR-TEM was performed at a Zeiss Libra 120 operating at 120 keV and a JEOL JEM- 2200FS operating at 200 keV, respectively. For material characterization, two distinct sample preparation techniques were applied. On the one hand, a drop of a diluted dispersion of magnetic nanoparticles extracted from the hydrogel was placed on a Formvar coated copper grid and left to dry on a filter paper. On the other hand, the grid was dipped inside the hydrogel matrix and aliquots were blotted using a filter paper. For microtome cut preparation, the ethanol dehydrated ferrogel samples were embedded in LR white Resin (Medium Grade) and cut with a Leica EM Trim.

*Analytical Methods—Electron Holography*: Off-axis electron holography of the magnetite chains grown in gelatin (Figure 2) was performed with a FEI Tecnai F20/Cs-corrected TEM operating at 200 kV acceleration voltage equipped with a Möllenstedt biprism and a 2k Slow-Scan CCD-camera (Gatan, USA) using the so-called pseudo-Lorentz mode. The separation of magnetic and electrostatic phase contribution was done by flipping the sample. For the nongelatin specimen (Figure 4), the measurement was performed using a Philips CM200FEG ST/LL TEM (FEI, Eindhoven, Netherlands) instrument operating at 200 kV acceleration voltage with a Möllenstedt biprism and a Multiscan 600HP 1k slow-scan CCD camera using the so-called Lorentz mode. The separation of magnetic and electrostatic phase contribution was done by magnetizing the sample with the field of the objective lens. Analysis of the TEM images was done with the software Digital Micrograph (Gatan, USA). Magnetostatic simulations with the magnetization as free parameter have been employed to determine the latter from the holographic data by minimizing the difference between experiment and simulation. Here, the shapes of the particles have been modeled from experimental data and we assumed a homogeneous polarization throughout the particle. Subsequently, we solved the well-known system of magnetostatic equations containing the scalar magnetic potential employing periodic boundary conditions at the edges of a sufficiently large supercell containing the particle chain.

*Analytical Methods—Tomography (Acquisition)*: For Acquisition of the Bright-field (BF)TEM tilt series, TEM specimen were investigated with the above mentioned Philips CM200FEG ST/LL TEM using a dedicated electron tomography holder (model 2040, Fischione Instruments). The tilt series covering an angular range from −78° to +66° in 3° steps was acquired in BFTEM imaging mode using a 20 μm contrast aperture inserted in the back focal plane of the objective lens, and recorded on a slow-scan CCD camera using a magnification that corresponds to 0.57 nm pixel size related to the object plane.

*Analytical Methods—Tomographic Reconstruction*: Before the actual tilt series alignment, the negative natural logarithm of the BFTEM images was computed to obtain the projected attenuation coefficient of the magnetite chains based on scattering absorption by the objective aperture. The tilt series were then pre-aligned by cross-correlation to correct for coarse displacements between successive projections. To perform the exact final alignment, i.e., the accurate determination of the tilt axis and correction for subpixel displacement, we have used an in-house-implemented center-of-mass method for correction of displacements perpendicular to the tilt axis in the sinograms, and the common line approach for the displacement correction parallel to the tilt axis. The 3D reconstruction of the aligned tilt series was carried out by weighted simultaneous iteration reconstruction technique (W-SIRT), a technique that uses a weighted back-projection at each iteration step. In our specific case, three iteration steps were sufficient to achieve a sound trade-off between high lateral resolution and low amplification of artifacts, e.g., due to diffraction contrast in the BFTEM images and incomplete tilt range ("missing wedge" artefacts).

## Supporting Information

Supporting Information is available from the Wiley Online Library or from the author.

## Acknowledgements

S.S. and M.S. contributed equally to this work. The authors thank Dr. Marc Widdrat for performing the synchrotron XRD measurements and Aleksej Laptev for SQUID measurements and Dr. Paul Simon for support and fruitful discussions. Last but not least, the authors greatly acknowledge the fundamental support by all members of the former Triebenberg Laboratory TU Dresden, especially Prof. Hannes Lichte. M.S., D.F., and H.C. thank the DFG for financial support within SPP 1569. E.S. and H.C. thank the DFG for support within the SFB 1214. E.S. thanks the Zukunftskolleg at the University of Konstanz for financial support. S.S., D.W., and A.L. have received funding from the European Research Council (ERC) under the European Union's Horizon 2020 research and innovation program (grant agreement No. 715620).

## Conflict of Interest

The authors declare no conflict of interest.

[1] E. Alphandéry, A. T. Ngo, C. Lefèvre, I. Lisiecki, L. F. Wu, M. P. Pileni, *J. Phys. Chem. C* **2008**, *112*, 12304.

[2] a) D. Faivre, D. Schüler, *Chem. Rev.* **2008**, *108*, 4875; b) I. Orue, L. Marcano, P. Bender, A. García-Prieto, S. Valencia, M. A. Mawass, D. Gil-Cartón, D. Alba Venero, D. Honecker, A. García-Arribas, L. Fernández Barquín, A. Muela, M. L. Fdez-Gubieda, *Nanoscale* **2018**, *10*, 7407.

[3] a) S. R. Mishra, M. D. Dickey, O. D. Velev, J. B. Tracy, *Nanoscale* **2016**, *8*, 1309; b) A. G. Meyra, G. J. Zarragoicoechea, V. A. Kuz, *Phys. Chem. Chem. Phys.* **2016**, *18*, 12768; c) X. Jiang, J. Feng, L. Huang, Y. Wu, B. Su, W. Yang, L. Mai, L. Jiang, *Adv. Mater.* **2016**, *28*, 6952.

[4] R. Wacker, B. Ceyhan, P. Alhorn, D. Schueler, C. Lang, C. M. Niemeyer, *Biochem. Biophys. Res. Commun.* **2007**, *357*, 391.

[5] B. Ceyhan, P. Alhorn, C. Lang, D. Schüler, C. M. Niemeyer, *Small* **2006**, *2*, 1251.

[6] T. Matsunaga, M. Takahashi, T. Yoshino, M. Kuhara, H. Takeyama, *Biochem. Biophys. Res. Commun.* **2006**, *350*, 1019.

[7] T. Matsunaga, K. Maruyama, H. Takeyama, T. Katoh, *Biosens. Bioelectron.* **2007**, *22*, 2315.

[8] E. Bäuerlein, D. Schüler, R. Reszka, S. Päuser, *US patent 6,251,365 B1*, **2001**.

[9] R. Hergt, R. Hiergeist, M. Zeisberger, D. Schüler, U. Heyen, I. Hilger, W. A. Kaiser, *J. Magn. Magn. Mater.* **2005**, *293*, 80.

[10] a) T. Matsunaga, A. Arakaki, in *Magnetoreception and Magnetosomes in Bacteria* (Ed: D. Schüler), Vol. *3*, Springer, Berlin Heidelberg **2007**, p. 227; b) T. Matsunaga, T. Suzuki, M. Tanaka, A. Arakaki, *Trends Biotechnol.* **2007**, *25*, 182.

[11] a) J. J. M. Lenders, C. L. Altan, P. H. H. Bomans, A. Arakaki, S. Bucak, G. de With, N. A. J. M. Sommerdijk, *Cryst. Growth Des.* **2014**, *14*, 5561; b) J. Liu, Z. Sun, Y. Deng, Y. Zou, C. Li, X. Guo, L. Xiong, Y. Gao, F. Li, D. Zhao, *Angew. Chem., Int. Ed.* **2009**, *48*, 5875; c) T. Prozorov, P. Palo, L. Wang, M. Nilsen-Hamilton, D. Jones, D. Orr, S. K. Mallapragada, B. Narasimhan, P. C. Canfield, R. Prozorov, *ACS Nano* **2007**, *1*, 228; d) A.-W. Xu, Y. Ma, H. Colfen, *J. Mater. Chem.* **2007**, *17*, 415.

[12] a) A. Arakaki, F. Masuda, Y. Amemiya, T. Tanaka, T. Matsunaga, *J. Colloid Interface Sci.* **2010**, *343*, 65; b) J. Baumgartner, M. Antonietta Carillo, K. M. Eckes, P. Werner, D. Faivre, *Langmuir* **2014**, *30*, 2129; c) T. Prozorov, S. K. Mallapragada, B. Narasimhan, L. Wang, P. Palo, M. Nilsen-Hamilton, T. J. Williams, D. A. Bazylinski, R. Prozorov, P. C. Canfield, *Adv. Funct. Mater.* **2007**, *17*, 951; d) V. Reichel, A. Kovacs, M. Kumari, E. Bereczk-Tompa, E. Schneck, P. Diehle, M. Posfai, A. M. Hirt, M. Duchamp, R. E. Dunin-Borkowski, D. Faivre, *Sci. Rep.* **2017**, *7*, 8.

[13] a) M. Helminger, B. Wu, T. Kollmann, D. Benke, D. Schwahn, V. Pipich, D. Faivre, D. Zahn, H. Cölfen, *Adv. Funct. Mater.* **2014**, *24*, 3187; b) M. Siglreitmeier, B. H. Wu, T. Kollmann, M. Neubauer, G. Nagy, D. Schwahn, V. Pipich, D. Faivre, D. Zahn, A. Fery, H. Colfen, *Beilstein J. Nanotechnol.* **2015**, *6*, 134.

[14] B. Wu, M. Siglreitmeier, C. Debus, D. Schwahn, H. Cölfen, V. Pipich, *Macromol. Biosci.* **2018**, *18*, 1800018.

[15] T. Sugimoto, E. Matijević, *J. Colloid Interface Sci.* **1980**, *74*, 227.

[16] a) K. Butter, P. H. H. Bomans, P. M. Frederik, G. J. Vroege, A. P. Philipse, *Nat. Mater.* **2003**, *2*, 88; b) M. Klokkenburg, C. Vonk, E. M. Claesson, J. D. Meeldijk, B. H. Erné, A. P. Philipse, *J. Am. Chem. Soc.* **2004**, *126*, 16706; c) E. V. Sturm, H. Colfen, *Chem. Soc. Rev.* **2016**, *45*, 5821; d) P. I. C. Teixeira, J. M. Tavares, M. M. Telo da Gama, *J. Phys.: Condens. Matter* **2000**, *12*, R411.

[17] C. L. Altan, J. J. M. Lenders, P. H. H. Bomans, G. de With, H. Friedrich, S. Bucak, N. A. J. M. Sommerdijk, *Chem.- Eur. J.* **2015**, *21*, 6150.

[18] R. Kniep, P. Simon, E. Rosseeva, *Cryst. Res. Technol.* **2014**, *49*, 4.

[19] E. T. Simpson, T. Kasama, M. Pósfai, P. R. Buseck, R. J. Harrison, R. E. Dunin-Borkowski, *J. Phys.: Conf. Ser.* **2005**, *17*, 108.

[20] a) L. Bergstrom, E. V. Sturm Nee Rosseeva, G. Salazar-Alvarez, H. Coelfen, *Acc. Chem. Res.* **2015**, *48*, 1391; b) E. Sturm, H. Cölfen, *Crystals* **2017**, *7*, 207.

[21] a) R. E. Dunin-Borkowski, M. R. McCartney, R. B. Frankel, D. A. Bazylinski, M. Pósfai, P. R. Buseck, *Science* **1998**, *282*, 1868; b) H. Lichte, F. Borrnert, A. Lenk, A. Lubk, F. Roder, J. Sickmann, S. Sturm, K. Vogel, D. Wolf, *Ultramicroscopy* **2013**, *134*, 126; c) P. A. Midgley, *Micron* **2001**, *32*, 167; d) J. M. Thomas, E. T. Simpson, T. Kasama, R. E. Dunin-Borkowski, *Acc. Chem. Res.* **2008**, *41*, 665; e) F. Röder, K. Vogel, D. Wolf, O. Hellwig, S. H. Wee, S. Wicht, B. Rellinghaus, *Ultramicroscopy* **2017**, *176*, 177.

[22] L. M. Peng, G. Ren, S. L. Dudarev, M. J. Whelan, *Acta Crystallogr., Sect. A: Found. Crystallogr.* **1996**, *52*, 257.

[23] O. Ozdemir, D. J. Dunlop, *Earth Planet. Sci. Lett.* **1999**, *165*, 229.

[24] Z. Nedelkoski, D. Kepaptsoglou, L. Lari, T. L. Wen, R. A. Booth, S. D. Oberdick, P. L. Galindo, Q. M. Ramasse, R. F. L. Evans, S. Majetich, V. K. Lazarov, *Sci. Rep.* **2017**, *7*, 8.

[25] J. Li, K. Ge, Y. Pan, W. Williams, Q. Liu, H. Qin, *Geochem., Geophys., Geosyst.* **2013**, *14*, 3887.